\makeatletter
\@ifundefined{@parse@version@dash}{%
\def\@parse@version#1{\@parse@version@0#1}
\def\@parse@version@#1/#2/#3#4#5\@nil{%
\@parse@version@dash#1-#2-#3#4\@nil}
\def\@parse@version@dash#1-#2-#3#4#5\@nil{%
  \if\relax#2\relax\else#1\fi#2#3#4 }
}{}
\makeatother

\documentclass[%
 aip,
 amsmath,amssymb,
 reprint,%
]{revtex4-1}

\usepackage{graphicx}
\usepackage{dcolumn}
\usepackage{bm}

\usepackage[utf8]{inputenc}
\usepackage[T1]{fontenc}
\usepackage{mathptmx}
\usepackage{etoolbox}
\usepackage{xcolor}
\def\dd{\mbox{d}}

\makeatletter
\def\@email#1#2{%
 \endgroup
 \patchcmd{\titleblock@produce}
  {\frontmatter@RRAPformat}
  {\frontmatter@RRAPformat{\produce@RRAP{*#1\href{mailto:#2}{#2}}}\frontmatter@RRAPformat}
  {}{}
}%
\makeatother
\begin{document}

\preprint{AIP/123-QED}

\title[Modeling the neurobiology of drug addiction]{A mathematical model of reward-mediated learning in drug addiction}
\author{Tom Chou}%
\affiliation{Department of Computational Medicine, UCLA, Los Angeles, CA 90095-1766, USA} 
\altaffiliation[Also at ]{Department of Mathematics, UCLA, Los Angeles, CA 90095-1555, USA}
\author{Maria R. D'Orsogna}%
 \email{dorsogna@csun.edu}
\affiliation{Department of Mathematics, California State University at Northridge, Los Angeles, CA 91130-8313, USA
}%
\altaffiliation[Also at ]{Department of Computational Medicine, UCLA, Los Angeles, CA 90095-1766, USA} 

\date{\today}

\begin{abstract}
Substances of abuse are known to activate and disrupt neuronal
circuits in the brain reward system.  We propose a simple and easily
interpretable dynamical systems model to describe the neurobiology of
drug addiction that incorporates the psychiatric concepts of reward
prediction error (RPE), drug-induced incentive salience (IST), and
opponent process theory (OPT).  Drug-induced dopamine releases
activate a biphasic reward response with pleasurable, positive
``a-processes'' (euphoria, rush) followed by unpleasant, negative
``b-processes'' (cravings, withdrawal).  Neuroadaptive processes
triggered by successive intakes enhance the negative component of the
reward response, which the user compensates for by increasing drug
dose and/or intake frequency.  This positive feedback between
physiological changes and drug self-administration leads to
habituation, tolerance and eventually to full addiction.  Our model
gives rise to qualitatively different pathways to addiction that can
represent a diverse set of user profiles (genetics, age) and drug
potencies.  We find that users who have, or neuroadaptively
develop, a strong b-process response to drug consumption are most at risk for addiction.
Finally, we include possible mechanisms to mitigate
withdrawal symptoms, such as through the use of methadone or other
auxiliary drugs used in detoxification.
\end{abstract}

\maketitle

\begin{quotation}
Drug abuse has been dramatically increasing worldwide over the last
twenty years.  Despite attempts to implement effective prevention
programs, treatment options, and legislation, drug poisoning remains a
leading cause of injury-related death in the United States, with a
record 100,000 fatal overdoses recorded in 2020.  Understanding how an
addiction to illicit substances develops is of crucial importance in
trying to develop clinical, pharmaceutical, or behavioral
interventions.  The neurobiological basis of drug addiction is
centered on disruptions to the dopamine system in the brain reward
pathway of users, which lead to neuroadaptive changes and the need for
larger or more frequent intakes to avoid withdrawal symptoms.  Despite
the many qualitative descriptions of the pathway to addition, a
concise mathematical representation of the process is still lacking.
We propose a unified, easily interpretable dynamical systems model
that includes the concepts of reward prediction error (RPE),
drug-induced incentive salience (IST) and opponent process theory
(OPT).  Specifically, we introduce a time-dependent reward function
associated with each drug intake. Physiological parameters evolve
through neuroadaptation, consistently with OPT, while user-regulated
drug intake is dependent on the most recent reward prediction,
consistent with RPE. Our model yields different distinct stages of the
addiction process that are cycled through via a dynamical recursion.
Individual-specific parameters may be tuned to represent different
drug potencies, age, or genetic predispositions. Rich features emerge,
such as monotonically convergent or damped oscillatory ("yo-yo")
progression towards full addiction.  Finally, our model can be used to
explore detoxification strategies.
\end{quotation}

\section{\label{sec:level1}Introduction}

Despite decades of medical, political, and legal efforts, substance
abuse remains a major issue worldwide.  The annual number of overdose
deaths in the United States has risen from about 20,000 in 2000 to
over 70,000 in 2019 \cite{Ahmad2020}, resulting in the highest drug
mortality rate in the world at an economic cost of at least $\$ 740$
billion USD per year \cite{NIDA2018}.

Our understanding of addiction, why and how it emerges, is still
incomplete although several mechanisms of action have been
identified \cite{Redish2008, Self1995} and
modeled \cite{GUTKIN2013,GUTKIN2014}. Addictive substances hijack the
mesocorticolimbic pathways which govern our response to primary
rewards such as food, drink, and sex. Under normal conditions, primary
rewards increase levels of dopamine, the main neurotransmitter in the
brain reward system.  Dopamine-strengthened neuronal connections
encode information on the reward and its
utility~\cite{Berridge2007,Volkow2009}, while its release in the
mesocorticolimbic pathways regulates incentive salience, the want and
seeking of rewards ~\cite{Berridge2003,Jones2005}.  To optimize future
responses, dopaminergic neurons respond differently to rewards that
deviate from expectations~\cite{Hollerman1998,Cohen2012,Hu2016}.  The
reward prediction error (RPE) quantifies the discrepancy between a
reward and its prediction and plays a major role in learning: neural
activity increases if the reward is greater than expected (positive
RPE) and decreases otherwise (negative
RPE) \cite{Oyama2010,Schultz2006, Schultz2016}. The RPE embodies
reinforcement learning, a key concept in psychology that has been
modeled and applied to many contexts, including drug
addiction \cite{GUTKIN2013}.

The effects of addictive drugs on the brain are similar to that of
primary rewards; drugs however amplify desires in abnormal ways.
Viewed as rewards, cocaine, amphetamines, and morphine act faster and
increase dopamine levels two to ten times more than food or
sex \cite{DiChiara1988, Kelley2004,Blum2012}, exaggerating the brain's
response to any drug-related cue. The operational mechanisms of each
drug type may be different, for example cocaine blocks the reuptake of
dopamine, whereas heroin binds to mu-opioid receptors which directly
stimulate the release of dopamine.  Other molecular targets of drugs
of abuse include the neurotransmitters endorphin and enkephalin
(particularly in the case of prescription opioids) and norepinephrine
and glutamate \cite{Cosgrove2010,Martinez2010}. Signaling between
different neurotransmitter types frequently leads to secondary
effects.  Whether directly or indirectly activated the most common
feature of drug intake is a dramatic increase in dopamine signaling in
the nucleus accumbens (NAc) \cite{Sulzer2011}, which is the process we
will focus on in our modeling.

Incentive sensitization theory (IST) formalizes the concepts
illustrated above \cite{Robinson1993}. Another relevant psychological
concept is opponent process theory (OPT) whereby every emotional
experience, pleasant or unpleasant, is followed by a counteracting
response to restore homeostasis.  Within OPT, the consumption of drugs
induces an ``a-process,'' marked by euphoria, rush, and pleasure,
later compensated by a ``b-process'' marked by withdrawal symptoms,
and craving \cite{Koob2001,Koob2008}. For beginning users, the
pleasant a-process is more intense and lasts longer than the
unpleasant b-process. Continued use leads to neuroadaptation, with the
b-process appearing earlier and lasting longer. Tolerance and
dependence set in \cite{George2017,Werner2019} as drug consumption
becomes predominantly unpleasant \cite{Robinson2003}.  Examples of
drug-dependent neuroadaptation include the reduction of postsynaptic
D$_2$ dopamine receptors~\cite{Seger2010}, neuronal
axotomy~\cite{Seger2010_2}, decreased dopamine neuron
firing \cite{Diana2011}, increases in the number of AMPA
receptors \cite{Ungless2001}, and activation of D1-like
receptors \cite{Dong2004}. Central among the brain tissues responding
to drug use is the ventral tegmental area (VTA) whose dopaminergic
neurons project to the nucleus accumbens (NAc) shell and to the
ventral pallidum (VP), two of the brain's pleasure centers associated
with the a-process. The neurobiological source of the b-process has
been identified with the subsequent activation of several stress
circuits controlled by the extended amygdala and the hypothalamus,
disrupting the release of stress related hormones or peptides such as
CRH, norepinephrine, dynorphin, hypocretin, leading to aversive
feelings \cite{Heinrichs1995,Koob2001,Koob2015,Shippenberg2017}.

How drugs impact the brain reward system has been mathematically
studied using dynamical
systems\,\cite{CHANGEUX,book,GUTKIN2013,Grasman2016,Radulescu2017,Duncan2019},
real-time neural networks\,\cite{Grossberg1987,Zhang2009},
temporal-difference reinforcement learning\,\cite{Redish2004}, and
model-free learning models\,\cite{huys2014}. While these models
explain certain observed features of the addiction process, a simpler,
yet explicit quantitative framework that unifies concepts from RPE,
IST, OPT, and allostasis is still lacking. Here, we construct and
analyze a proof-of-principle mathematical model of the onset of drug
addiction, resolved at the individual drug intake time scale.
Neuroadaptation is represented by changes in physiological parameters
consistent with RPE and OPT, informing changes to user behavior.
These changes induce further neuroadaptation, creating a feedback loop
that may lead to full addiction.  
We introduce measures to
  quantify the overall reward resulting from a single drug intake and
  for the reward prediction error; addiction is mathematically defined
  as the state in which the overall reward is negative and
  the reward prediction error is below a given threshold.  These
  formulations allow us to predict the unfolding of the addiction
  process depending on the specific physiology and neuroadaptive
  profile of the user. Specifically, we find that, given the same
  drug, users who are more sensitive to neuroadaptive changes in the
  b-process (or who have an initially elevated b-process response) are
  the ones whose progression to addiction is faster than those who are
  less reactive. These more resilient users may also display 
  reward prediction errors that oscillate in value with each drug intake
  before permanently crossing the threshold to addiction  (``yo-yo''  dynamics).

\section*{Mathematical Model}
\label{materials}

\begin{figure}[t]
  \centering{\includegraphics[width=3.4in]{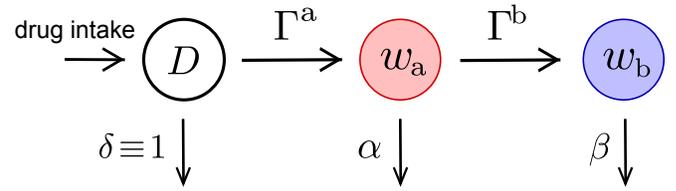}}
\caption{Schematic of a- and b-processes. Drug use 
activates the dopaminergic neurons which in turn activate the hedonic
  hotspots in the nucleus accumbens that mediate the pleasurable
  ``a-processes'' $w_{\rm a}$, leading to euphoria and bliss.
  Unpleasurable ``b-processes'' $w_{\rm b}$ may follow, accompanied by
  cravings and withdrawal symptoms. The relative magnitude of the two
  $w_{\rm a,b}$ experiences may vary among individuals and may depend
  on the stage of addiction. Dopamine-induced activity is modeled as
  $D(t) = \Delta e^{-\delta t}$ where $\Delta$ is a proxy for drug
  dosing and $\delta$ its typical degradation rate. The overall
  a-process is activated by $D(t)$ via the prefactor $\Gamma^{\rm a}$,
  whereas the overall b-process is activated by the a-process via the
  prefactor $\Gamma^{\rm b}$. The activity of the a- and b-processes
  decay with rates $\alpha$ and $\beta$, respectively.}
\label{FIG0}
\end{figure}

\begin{figure*}[t]
  \centering
\includegraphics[width=4.8in]{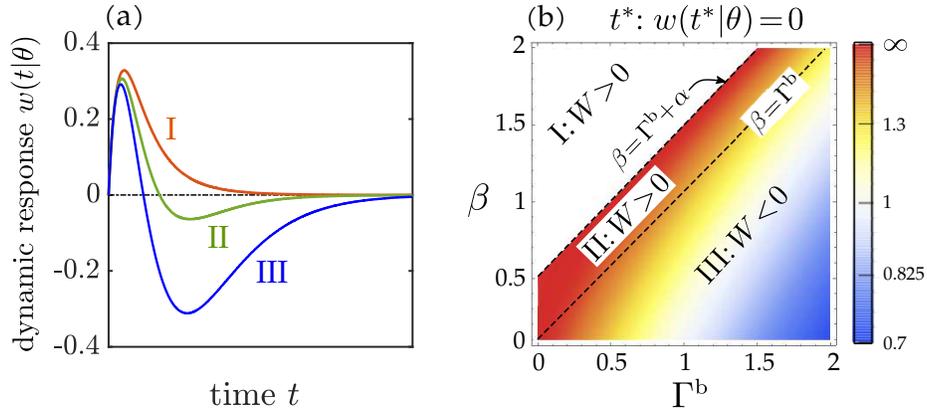}
\caption{(a)
  Three examples of time-dependent response $w(t\vert \theta)$
  associated with a single, isolated dopamine hit. We fix $\theta
  = \{\Delta=1,\alpha=0.5, \Gamma^{\rm a}=1,\beta, \Gamma^{\rm
  b}=0.8\}$ and plot Eq.~\ref{sol_w} for three different values of
  $\beta$ showing a response that is always positive (${\rm I}$:
  $\beta=1.5$, orange), a response that can become negative (${\rm
  II}$: $\beta=0.9$, green), and one with a negative total reward
  (${\rm III}$: $\beta=0.45$, blue). (b) Density plot representing the
  values of the time $t^*$ associated with the solution to the
  transcendental equation $w(t^*\vert \theta)=0$ as a function of $\beta$
  and $\Gamma^{\rm b}$ for $\alpha=0.5$. The white parameter region
  does not admit a finite solution to $t^*$ ($w(t|\theta)$ is always
  positive). }
\label{FIG2}
\end{figure*}

\subsection*{Dopamine release}
\label{dopamine}
We begin by describing the time-dependent activity $D(t)$
(\textit{e.g.}, firing rate) of dopaminergic neurons in the reward
system in response to the dopamine release that follows a single,
initial drug intake.  Other rewards such as food, sex, etc. also
stimulate dopamine release, however, it is known that drug-induced
dopamine release is an order of magnitude larger than what stimulated
by ``natural'' rewards \cite{DiChiara1988}. Experimental measurements
show a rapid rise in activity within a few minutes of intravenous drug
administration \cite{Volkow2000,Fowler2008}, followed by an
exponential decay over one to five hours \cite{DiChiara1988}. We
propose
\begin{equation}
\label{activity}
    D(t) =  \Delta e^{-\delta t},
\end{equation}
where $\Delta$ is the magnitude of the dopamine response and
$1/\delta$ is the effective dopamine residence time which includes the
clearance time of dopamine-stimulating drugs; typically $\delta \sim
0.2-1$ hr$^{-1}$. For simplicity we measure time in units of $\delta$,
rescale $t' \to \delta t$, drop the prime notation, and set $D(t) =
\Delta e^{-t}$.  Since dopamine release is triggered by drug
intake we henceforth use $\Delta$ as a proxy for drug dosage. Although
more complex pharmacokinetic models have been developed to connect
drug dose to dopamine activity \cite{book_pharm}, the time dependence
of dopamine activity qualitatively resembles a decaying exponential
except at very short times.

\subsection*{Single-dose drug-induced a- and b-processes}

According to OPT and as described above, drug-induced dopamine
activity $D(t)$ induces a pleasurable a-process, $w_{\rm a}(t)$, which
in turn activates an unpleasant b-process, $w_{\rm b}(t)$ (see
Fig.\,\ref{FIG0}).  We propose a deterministic model for $w_{\rm
a,b}(t)$ that incorporates simple integrate-and-fire dynamics

\begin{eqnarray}
\frac{\dd w_{\rm a}(t)}{\dd t} &=& -\alpha w_{\rm a}(t) + \Gamma^{\rm a}D(t) 
 \label{wa} \\
\frac{\dd w_{\rm b}(t)}{\dd t} &=&  -\beta w_{\rm b}(t)
- \Gamma^{\rm b} w_{\rm a}(t),
 \label{wb}
\end{eqnarray}
where $\Gamma^{\rm a}$ and $\Gamma^{\rm b}$ represent the coupling of
$D(t)$ to $w_{\rm a}(t)$, and of $w_{\rm a}(t)$ to $w_{\rm b}(t)$,
respectively.  The intrinsic decay rates of the a- and b-processes are
denoted $\alpha$ and $\beta$. The effects of intermittent natural
rewards that induce dopamine release can be incorporated by including
an extra source to $w_{\rm a}(t)$ in the form of a periodic or a
randomly fluctuating term. These non-drug terms would be much smaller
in magnitude than the drug source $D(t)$, since drug induced stimuli
are much larger than non-drug ones \cite{DiChiara1988}.  The periodic
part may represent, say, eating at regular intervals, whereas the
fluctuating part might describe all other non-drug, pleasurable
experiences that occur at random times. Thus, a stochastic model might
yield a more complete description of the brain reward system and its
many inputs but we shall limit this study to the deterministic
response from well-defined drug intakes as presented in
Eqs.\,\ref{wa}-\ref{wb}.

The $w_{\rm a,b}(t)$ processes generate the brain reward system's
perception of the drug. While further complex processing and filtering
of $w_{\rm a,b}(t)$ may be at play, we assume they are summed to yield
the dynamic, time-dependent response $w(t\vert \theta) = w_{\rm
a}(t)+w_{\rm b}(t)$ where $\theta =
\{\Delta, \alpha, \Gamma^{\rm a}, \beta, \Gamma^{\rm b}\}$ are the
parameters associated with the reward perception process.  Upon
solving Eqs.\,\ref{wa} and \ref{wb}, we find the dynamic response $w(t
\vert \theta)$ following a single, isolated dopamine release and/or
drug intake

\begin{eqnarray}
\nonumber
w(t\vert \theta) 
& =&  \frac{\Gamma^{\rm a}\Delta}{\alpha - 1}
\bigg[\left(1-\frac{\Gamma^{\rm b}}{\beta-1}\right)e^{-t}
-\left(1-\frac{\Gamma^{\rm b}}{\beta-\alpha}\right) e^{-\alpha t} \\
& & \hspace{6mm}-\left(\frac{\Gamma^{\rm b}}{\beta - \alpha}
  -\frac{\Gamma^{\rm b}}{\beta-1}\right) e^{-\beta t}\bigg].
  \label{sol_w}
\end{eqnarray}
The time-integrated net response 

\begin{equation}
\label{total_W}
W(\theta) = \int_{0}^{\infty} 
w(t\vert \theta) \dd t = \frac{\Gamma^{\rm
    a}\Delta}{\alpha}\left(1-\frac{\Gamma^{\rm b}}{\beta}\right),
\end{equation}
associated with a single, isolated dopamine release and/or drug intake
can be interpreted as a memory of the experience, and can be used as a
benchmark for future decision-making.
Note that the amplitude factor $\Gamma^{\rm a}$ in Eqs.~\ref{sol_w}
and \ref{total_W} adjusts the ``hedonic'' scale of $w(t\vert \theta)$
and $W(\theta)$.

\begin{figure}[t]
  \centering
  \includegraphics[width=3.3in]{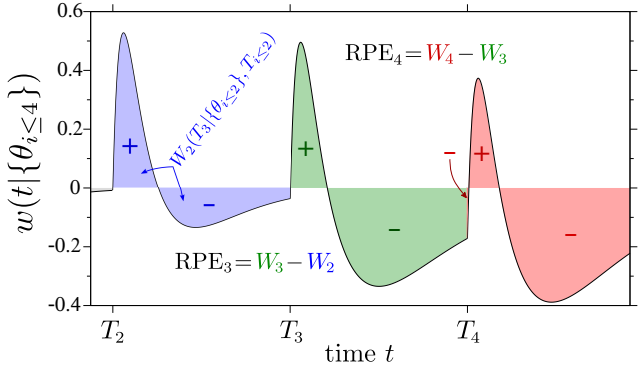}
  \caption{Time-dependent response $w(t)$ resulting from multiple drug
    intakes at times $T_1=0$ (not shown), $T_2$, $T_3$, and
    $T_{4}$. Each dopamine release elicits an a- and b-process
    response which can be concatenated (Eq.\,\ref{w_sum}). The net
    reward $W_{k}$ associated with dose $k$ is defined as the integral
    of $w(t|\{\theta_{i\leq k}\})$ from time $T_{k}$ to $T_{k+1}$ and
    may depend on $\theta_{i\leq k}$ since the a- and b-processes
    triggered by previous drug intakes may not have fully
    dissipated. For small $\beta_{i}$, b-processes relax slowly making
    the response to appear to reach a lower homeostatic value.  The
    small $\beta$ regime resembles the allostatic effect on time
    scales $\lesssim 1/\beta$. In this limit, repeated drug doses
    successively drive the reward response negative pushing the user
    to experience increasingly intense withdrawal symptoms. In these
    plots, $\Delta=1, \alpha=0.2, \Gamma^{\rm a}=1, \Gamma^{\rm
    b}=0.2$ and $\beta_1=0.3$, $\beta_2=0.2$, $\beta_{3}=0.1$, and
    $\beta_4=0.1$ for intakes at $0, T, 2T$ and $3T$,
    respectively. The RPE is defined by the difference between two
    consecutive time-integrated responses $W_{k}-W_{k-1}$.}
\label{FIG3}
\end{figure}

In Figure \ref{FIG2}(a), we plot $w(t\vert\theta)$ for $\Delta=1,
\alpha=0.5, \Gamma^{\rm a}=1,\Gamma^{\rm b}=0.8$, and $\beta=1.5$
(orange curve ${\rm I}$), $\beta=0.9$ (green curve ${\rm II}$), and
$\beta=0.45$ (blue curve ${\rm III}$).  These representative response
curves $w(t\vert \theta)$ are: I, always positive; II, turning
negative with positive integral $W(\theta)>0$; and III, turning
negative with negative integral $W(\theta) <0$.  Type I responses are
typical of healthy, na\"ive users who for the most part experience
only the pleasurable a-process.  For smaller $\beta$, larger
$\Gamma^{\rm b}$, and/or larger $\alpha$ ($\Gamma^{\rm b} < \beta <
\Gamma^{\rm b}+\alpha$), $w(t\vert \theta)$ exhibits a type II
response which is negative at late times but yields a positive net
response $W(\theta)>0$.  For even smaller $\beta$ and/or larger
$\Gamma^{\rm b}$, the response is type III: the negative b-process
overtakes the a-process and the overall experience is negative with
$W(\theta)<0$.  Type II and type III responses are typical of moderate
and addicted users, respectively.  Figure\,\ref{FIG2}(b) shows the
density plot of the time $t^*$ when the dynamic response
$w(t^*\vert\theta)=0$, as a function of $\beta$ and $\Gamma^{\rm b}$ at
$\alpha=0.5$. For $\beta\geq \Gamma^{\rm b}+\alpha$ there are no
finite solutions $t^*$ to $w(t^*\vert \theta)=0$; in this regime the
dynamic response is always positive as represented by the type I curve
in Figure \ref{FIG2}(a).  For $\beta < \Gamma^{\rm b}+\alpha$, $t^*$
is positive and finite and decreases as $\beta$ decreases or
$\Gamma^{\rm b}$ increases, indicating a stronger overall
b-process. Examples are the type II and type III curves in Figure
\ref{FIG2}(a).

What we have described so far is a simple single-dose picture of the
reward response. In the next section we build on it to describe
addiction as a progression of multiple drug intakes that induce
neuroadaptive changes to the physiological parameters, $\beta$ and
$\Gamma^{\rm b}$, and behavioral changes to the user that shift the
net response from type I to type III.

\subsection*{Successive drug intakes}

We now consider successive drug intakes $i$ taken at times $T_{i}$ with
the first dose taken at $T_{1}=0$ and the most recent one at $T_{k}$. For
finite $T_{k}$, the total time-dependent response is a superposition
of the time-shifted responses in Eq.~\ref{sol_w} 
\begin{equation}
  w(t\vert \{\theta_{i\leq k}\}) = \sum_{i=1}^{k}
w(t - T_{i}\vert \theta_{i}), \,\,\, T_{1}\equiv 0, \,\, T_{k} < t < T_{k+1},
  \label{w_sum}
\end{equation}
where $\theta_i=\{\Delta_{i}, \alpha_i, \Gamma^{\rm
  a}_{i},\beta_{i},\Gamma^{\rm b}_{i}\}$
are the parameters of the system following intake $i$.  The doses
$\Delta_{i}$ and intake times $T_{i}$ are primarily
user-controlled. We assume the other parameters
$\{\alpha_i, \Gamma^{\rm a}_{i},\beta_{i},\Gamma^{\rm b}_{i}\}$ evolve
in a step-wise fashion due to dopamine-induced neuroadaptive changes,
such as long term potentiation or other long-lasting physiological,
tissue-level, or biochemical processes.  The total net response after
the last dose at time $T_{k}$ can be defined as an integral over
$w(t\vert \{\theta_{i\leq k}\})$ starting from $T_{k}$ until the
current time $t$. Thus, the net response associated with dose $k$ is
\begin{align}
  W_{k}(t\vert \{\theta_{i \leq k},T_{i\leq k}\}) & 
= \int_{T_{k}}^{t} w(t'\vert \{\theta_{i\leq k}\})\dd t',
\label{WT1}
\end{align}
where $T_{k} < t < T_{k+1}$ and $T_{k+1}$ is the time of the next
dose, if it occurs.  Using Eqs.~\ref{sol_w} and \ref{WT1}, we find

\begin{eqnarray}
\nonumber
  W_{k}(t\vert \{\theta_{i \leq k},T_{i\leq k}\}) &= &
  \sum_{i=1}^{k}C_{i}(e^{-(T_{k}-T_{i})}-e^{-(t-T_{i})}) + \\
&& \sum_{i=1}^{k}C^{\alpha}_{i}(e^{-\alpha_{i}(T_{k}-T_{i})}-e^{-\alpha_{i}(t-T_{i})}) +  \nonumber \\
&&  \sum_{i=1}^{k}C^{\beta}_{i}(e^{-\beta_{i}(T_{k}-T_{i})}-e^{-\beta_{i}(t-T_{i})}),
\label{WT2}
\end{eqnarray}
where 
\begin{eqnarray}
\nonumber
C_{i} &\equiv&  \frac{\Gamma^{\rm a}_{i}\Delta_{i}}{\alpha_{i}-1}\left(1-\frac{\Gamma^{\rm b}_{i}}{\beta_{i}-1}\right), \\
\label{parameters}
C_{i}^{\alpha} &\equiv&  \frac{\Gamma^{\rm a}_{i}\Delta_{i}}{\alpha_{i}-1} \frac{1}{\alpha_{i}}\left(\frac{\Gamma^{\rm b}_{i}}{\beta_{i}-\alpha_{i}}-1\right),\\
\nonumber
C_{i}^{\beta} &\equiv&  \frac{\Gamma^{\rm a}_{i}\Delta_{i}}{\alpha_{i}-1}
\frac{1}{\beta_{i}}\left(\frac{\Gamma^{\rm b}_{i}}{\beta_{i}-1}-\frac{\Gamma^{\rm b}_{i}}{\beta_{i}-\alpha_{i}}\right).
\label{C}
\end{eqnarray}
If drug intakes are well-separated ($T_{i+1}-T_{i}\to \infty$)
with no residual effects from previous doses, the net response
between $T_{k}$ and $T_{k+1}$ is $W_{k}(T_{k+1}\vert \{\theta_{i
  \leq k}, T_{i\leq k}\}) \to \Gamma^{\rm
  a}_{k}\Delta_{k}(1-\Gamma^{b}_{k}/\beta_{k})/\alpha_{k}$, the result
given in Eq.~\ref{total_W}.

\subsection*{Reward prediction error (RPE) and behavioral changes}

To construct the total time-dependent response for multiple drug
intakes $w(t\vert \{\theta_{i\leq k}\})$ in Eq.\,\ref{w_sum}, we must
describe the evolution of the user-controlled variables $\{\Delta_i,
T_i \}$ and of the neuroadaptive parameters $\{\alpha_i, \Gamma_i^{\rm
  a}, \beta_i, \Gamma_i^{\rm b} \}$ as a function of the number of
intakes $i$. In this section, we provide a mathematical description of
the RPE, the difference between the expected and received rewards
associated to each drug intake.  The RPE is a key component of
learning and decision-making; here it will be assumed to regulate the
specific decision of the user to change (or not) the next drug dose
$\Delta_{k+1}$ of intake $k+1$.

The expected response of a drug intake depends on the user's
prior history, experiences, and cues of upcoming rewards. The
expectation may be different from the actual, obtained response
leading to an error, the RPE. For simplicity, we represent the
RPE$_{k}$ following intake $k$, and just before intake $k+1$, as the
difference between the most recent net response $W_{k}$ and the 
prior one $W_{k-1}$
\begin{eqnarray}
\nonumber
    \text{RPE}_k &\equiv& W_{k}(T_{k+1}\vert \{\theta_{i\leq k},T_{i\leq k}\}) \\
&& -\gamma_{k-1}W_{k-1}(T_{k}\vert \{\theta_{i\leq k-1}, T_{i\leq k-1}\}) 
\label{rpe}
 -C_{k+1},
\end{eqnarray}
weighted by a factor $\gamma_{k-1}<1$ that discounts the previous net
response $W_{k-1}$ and that may incorporate memory effects. RPEs that
rely on responses associated with drug doses further in the past can
also be used to reflect longer memory of the
reward \cite{CHANGEUX}. The term $C_{k+1}$ is a history-independent
cue associated with the upcoming $k+1^{\rm th}$ intake. Examples of
cues include seeing or smelling the drug, or preparing for its
consumption. Without loss of generality, we assume $C_k =0$ by
shifting the baseline value of the RPE.  An example of a negative RPE
is shown in Fig.~\ref{FIG3}, where RPE$_{3} < 0$ indicating unmet
expectations from intake 3.  As defined in Eq.~\ref{rpe}, a positive
${\rm RPE}_k$ arises if $W_{k}>
\gamma_{k-1}W_{k-1}$, raising expectations for future intakes.  This
increased expectation may represent habituation, whereby continued use
generates a desire for greater net responses.  The value of RPE$_{k}$
will be used in the next section to determine if a behavioral change
-- another drug intake at time $T_{k+1}$ and/or a change in dose
$\Delta_{k+1}$ -- is elicited.

\begin{figure*}[t!]
  \centering
  \includegraphics[width=7in]{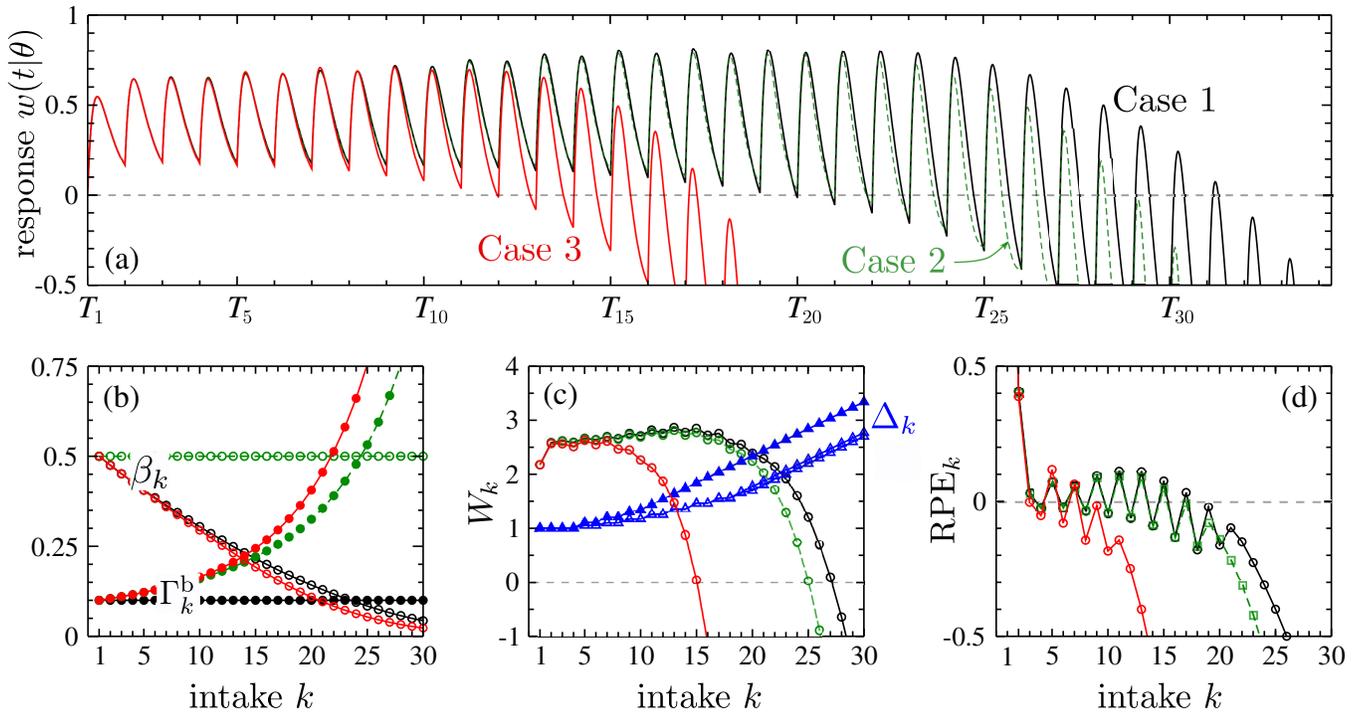}
  \caption{(a) The time-dependent response resulting from multiple
    drug intakes with varying $\Delta_k$ for three scenarios.  In Case
    1 (solid black curve), we set $B=0.05, G=0$ while in Case 2
    (dashed green curve), $B=0, G=0.05$. Finally, in Case 3 (solid red
    curve), $B=G=0.05$.  The interval between two consecutive intakes
    in these examples is $T=T_{k+1}-T_{k}=6$. The effects of
    neuroadaptation when both $\beta_{k}$ and $\Gamma_{k}^{\rm b}$
    evolve are synergistic as Case 3 leads to addiction after
    significantly fewer intakes.  (b) Evolution of the parameters
    $\beta_k$ (open circles, solid curves) and $\Gamma_k^{\rm b}$
    (filled circles, dashed curves) for Cases 1, 2, and 3. (c)
    Evolution of the integrated response and intake doses $\Delta_k$
    (blue triangles, Eq.~\ref{RECURSION_DELTA}) associated with intake
    $k$ for each of the three cases. The evolution of $\Delta_{k}$ is
    similar for Cases 1 and 2, while $\Delta_{k}$ for Case 3 rises
    faster and might describe a highly addictive drug that results in
    addiction after a smaller number of doses. The integrated
    responses $W_{k}$ become negative at about intake $k\approx 27,
    26$, and $16$ for Cases 1, 2, and 3, respectively.  (d) The
    RPE$_k$ as a function of the intake $k$ exhibit oscillations with
    increasing then decreasing amplitude before monotonically
    decreasing well below zero at intakes $k\approx 20, 18$, and 10
    for Cases 1,2, and 3, respectively. In all cases, the user
    experiences a ``yo-yo'' progression to addiction. Since in all
    cases, $W_{k}$ becomes negative after the RPE, addiction occurs
    when $W_{k}<0$ at intakes $k^{*} \approx 27, 26$, and 16,
    respectively.}
\label{FIG4}
\end{figure*}

\subsection*{Neuroadaptation and parameter changes}

In addition to $\{\Delta_, T_i \}$, the time-dependent response
$w(t\vert \{\theta_{i\leq k}\})$ also depends on the physiological
parameters $\{\alpha_i, \Gamma^{\rm a}_{i}, \beta_{i}, \Gamma^{\rm
b}_{i}\}$. Changes in these quantities can be driven by neuroadaptive
processes following each drug intake and can depend on the specific
characteristics (age, gender, constitution, genetic makeup) of each
user.  These neuroadaptive processes are complex and difficult to
model, so we simplify matters by assuming that
$\{\alpha_i, \Gamma^{\rm a}_{i}, \beta_{i}, \Gamma^{\rm b}_{i}\}$
change only in response to each drug-induced dopamine release
$\Delta_{i}$ at $T_{i}$.  To be consistent with OPT and observations,
neuroadaptative changes should increase the effects of the negative
b-process relative to those of the positive a-process as addiction
progresses. This can be achieved through a decrease in $\beta_{i}$
and/or an increase in $\Gamma^{\rm b}_{i}$.  In principle, changes in
$\alpha_{i}$ arising from tolerance (that shortens the ``high'' and
affects the relative strengths of the a- and b-processes) can also be
modeled, but since changes to $\Gamma^{\rm a}_{i}\Delta_i/(\alpha_{i}
- 1)$ only rescale $w(t)$, we fix $\alpha_{i} = \alpha$ and
$\Gamma^{\rm a}_{i}=\Gamma^{\rm a}$ to constant values. Thus, we let
$\Delta_{i}$ drive neuroadaptive changes in $\beta_{i+1}$ and
$\Gamma^{\rm b}_{i+1}$ according to the simplest rule consistent with
OPT:
\begin{equation}
\beta_{i+1} = \beta_{i}(1 - B\Delta_{i+1}), \quad
\Gamma_{i+1}^{\rm b} = \Gamma_{i}^{\rm b}(1 + G\Delta_{i+1}).
\label{RECURSION_BG}
\end{equation}
Here, $B$ and $G$ are parameter-change sensitivities that may depend
on $i$, $\beta_{i}$, $\Gamma_{i}^{\rm b}$, and $\Delta_i$, but that we
assume to be constant, with the caveat that $B$ is small enough that
for all values of $i$, $B \Delta_i < 1$. Equation \ref{RECURSION_BG}
implies that $\beta_{i}$ and $\Gamma_{i}^{\rm b}$ are represented by
piecewise constant values that change after each drug intake.  Note
that after a sufficient number of intakes $\beta_i$ becomes very small
and the negative response persists for a long time, yielding an
apparent ``allostatic'' state \cite{Mcewen1998}.  In the above
recursion Eq.\,\ref{RECURSION_BG}, the drug doses $\Delta_{i}$ may be
assumed fixed or may evolve according to models that involve the RPE.

We now incorporate the ingredients described above into a dynamical
model that generates trajectories to addiction. In this model, the
neuroadaptative evolution of the physiological parameters induces
changes to the reward responses, which in turn modify the RPE and lead
to user behavioral changes such as increases in drug dose or intake
frequency to boost the pleasurable a-process.  Despite these
user-controlled changes, the evolving neurophysiological parameters may
eventually lead to negative net responses and RPEs.  We thus define
addiction as a state marked by persistently negative RPE$_{i}<0$ and
negative net responses $W_{i}<0$ that arise for intakes at or greater
than a critical number $i \geq k^{*}$.

\subsection*{Evolution of intake doses}

We first consider the case where the intake times $T_{i} = (i-1)T$ are
perfectly periodic with interval $T$ and study the evolution of the
most recent dose $\Delta_k$ to the next one $\Delta_{k+1}$.  Although
more intense dopamine activity may be stimulated by a larger
$\Delta_{k+1}$ (according to Eq.\,\ref{sol_w} for well spaced
intakes), the resulting net response $W_{k+1}$ may not necessarily be
larger than $W_{k}$ since $W_{k+1}$ depends not only on dose, but also
on the neuroadaptive parameters $\{\alpha_{k+1},\Gamma^{\rm
a}_{k+1}, \beta_{k+1},\Gamma^{\rm b}_{k+1}\}$ over which the user has
no direct control. Thus, scenarios may arise in which although the
drug dose increases, the RPE remains negative and user expectations
are not met.  We assume that if the ${\rm RPE}>0$ the user will not
alter drug dose; however, if ${\rm RPE}<0$ the user will increase it.
To concretely model this behavior and allow variable $\Delta_{k}$, we
augment the recursion relations \ref{RECURSION_BG} as follows
\begin{equation}
\label{RECURSION_DELTA}
\Delta_{k+1} = \Delta_{k} + \sigma H({\rm RPE}_{k}), 
\qquad H(x) = \left\{\begin{array}{cc} 1 &  x \leq R_{\rm c} \\
\frac{x}{R_{\rm c}}  &   R_{\rm c}< x < 0 \\
 0 &  x \geq 0, \end{array} \right.
\end{equation}
where $\sigma$ is the maximal dose-change and $H(x)$ dictates how
doses increase as a function of RPE$_{k}$.  We choose the simple form
in Eq.~\ref{RECURSION_DELTA} representing a graded switching function
with threshold $R_{\rm c}/2$.  We use the representation of RPE$_k$
given in Eq.~\ref{rpe} in which for simplicity we set $\gamma_{k-1} =
1$ and $C_{k+1}=0$.  Finally, note that the argument of $H$ in
Eq.~\ref{RECURSION_DELTA}, ${\rm RPE}_{k}$, depends on the drug intake
period $T$ and the dose $\Delta_{k}$ through $W_k(T_{k+1}\vert
\{\theta_{i \leq k}, T_{i \leq k}\})$, which makes the evolution
Eq.~\ref{RECURSION_DELTA} non-linear.

In our model, changes to the neuroadaptive parameters
$\beta_{i+1}, \Gamma^{\rm b}_{i+1}$ at intake $i+1$ carry a linear
dependence on the dosage $\Delta_{i+1}$, according to
Eqs.\,\ref{RECURSION_BG}.  We adopted this choice for simplicity;
however more complex forms for the evolution of
$\beta_{i+1}, \Gamma^{\rm b}_{i+1}$ and $\Delta_{i+1}$ can be used to
study a wider range of scenarios.

 Specifically, the parameters
coefficients $B,G$ in Eqs.\,\ref{RECURSION_BG} could be modeled
to be functions of intake
number or time on drugs, through forms that depend on the genetics of age of the user.
Such refinements may be important
especially if one is interested in the long-term dynamics of drug
consumption, or in comparing responses among different user types.
For example, it is well known that drugs of abuse can significantly
impact the still-maturing, and thus vulnerable, adolescent brain and
cause severe, long-term damage \cite{Guerri2019}. Eqs.\,\ref{RECURSION_BG} can also be
modified to include saturation, or recovery of the baseline values of
$\beta_{i+1}, \Gamma^{\rm b}_{i+1}$ if the user stops using drugs.
Other nonlinearities may be introduced to represent distinct
neuroadaptive regimes. These could be stages of more (or less)
impactful changes once a given threshold of, say, drug dose,
cumulative drug dose, or reward value is reached \cite{Pando2021}.
These choices may lead to non-trivial dynamics involving
$\beta, \Gamma_{\rm b}, \Delta$ as well as the RPE, and possibly lead
to chaotic behaviors \cite{Li1975}, as proposed in the context of
alcohol addiction \cite{Hawkins1998, Skinner1989, Warren2003}.

\begin{figure*}[t]
  \centering
   \includegraphics[width=7in]{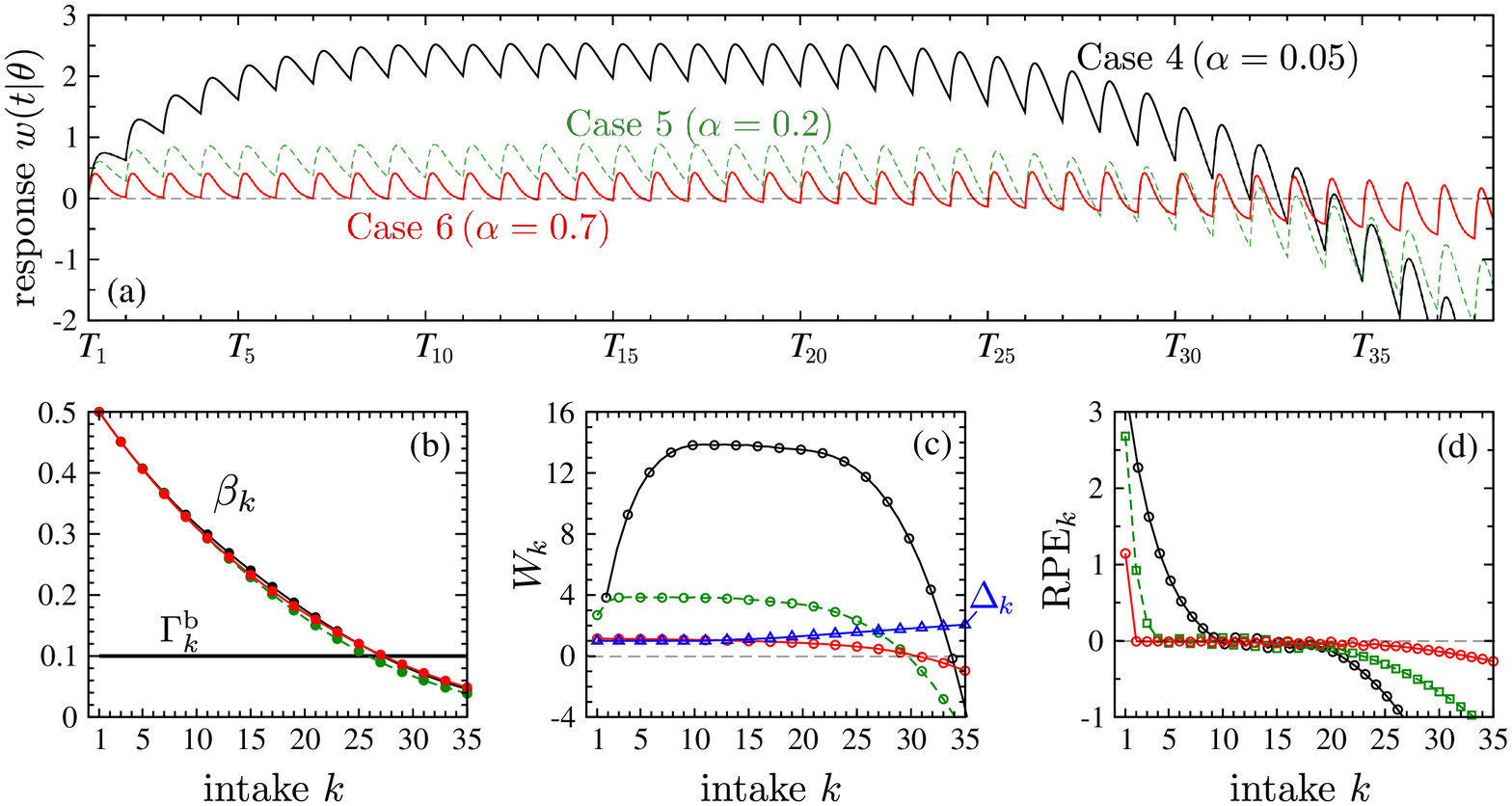}
 \caption{(a) The time-dependent response resulting from multiple drug
   intakes at times $T_{i} = (i-1)6$ with user-adjusted $\Delta_i$ for
   three additional scenarios corresponding to different durations
   $1/\alpha$ of the a-process.  In Case 4 (black solid curve), we
   keep $\beta_{1} = 0.5, \Gamma^{a} = 1, \Gamma^{b}_{1} = 0.1, \sigma
   = B=0.05, G=0$ but assume a long-lasting a-process by setting
   $\alpha = 0.05$.  In Cases 5 (green dashed curve) and 6 (solid red
   curve) we use $\alpha = 0.2$ and $\alpha = 0.7$, respectively.  The
   corresponding $\beta_{k}$ and $\Gamma_{k}^{\rm b}$ for these cases
   are nearly indistinguishable, as shown in (b). The corresponding
   doses (blue triangles) shown in (c) are also indistinguishable. The
   integrated responses $W_{k}$ for these three cases reach long-lived
   plateaus of different amplitudes.  The associated ${\rm RPE}$s
   are also qualitatively different, as shown in (d). In all cases,
   the ${\rm RPE}$s hover around small values for many intakes. Note
   while longer-lasting a-processes generate higher values of $W_{k}$,
   the corresponding RPEs decrease faster. Addiction in these three
   cases occur at $k^{*}\approx 33, 30$, and $31$ when $W_{k}<0$ since
   ${\rm RPE}_{k}<0$ occurs at $k \approx 11,5$ and $2$.}
\label{FIG5}
\end{figure*}

\section*{Results}
\label{rd}

We now study the effects of multiple intakes utilizing the full model
given by Eqs.\,\ref{WT1}--\ref{parameters} and
Eqs.\,\ref{rpe}--\ref{RECURSION_DELTA}.
We set $\alpha_i =0.3, \Gamma^{\rm a}_{i}=1, R_{\rm c}=-0.05,
\sigma=0.1$, 
initialize the system with $\{\beta_{1}, \Gamma^{\rm
b}_{1}, \Delta_{1}, {\rm RPE}_{1}\} = \{0.5, 0.1, 1, 0\}$.
Upon specifying $B,G$ we can find the first net response
$W_1(T_{2}\vert \{\theta_{1}, T_{1}\})$ per Eq.\,\ref{WT1}. We let the
second dose $\Delta_2=1$ and generate $\{\beta_{2}, \Gamma^{\rm
b}_{2}\}$ from Eq.~\ref{RECURSION_BG} and
$W_{2}(T_3 \vert \{\theta_{i \leq 2}, T_{i \leq 2}\})$ from
Eq.\,\ref{WT1}, yielding RPE$_{2}=W_2 (T_3 \vert \{\theta_{i \leq 2},
T_{i \leq 2}\}) -W_1(T_2 \vert \{\theta_{1}, T_{1}\})$. The next dose
$\Delta_3$ is then determined through Eq.~\ref{RECURSION_DELTA}, and
so on.  To illustrate responses to multiple fixed-period intakes, we
must specify the dimensionless time $T$ between drug intakes relative
to the drug-induced dopamine mean residence times.  In
Fig.~\ref{FIG4}, we assume the inter-intake period $T$ to be six times
the effective dopamine residence time $1/\delta$. Thus, if $\delta
\approx 0.25$/hr, daily drug dosing (once every 24hrs) corresponds to
$T=6$.
 
Figure~\ref{FIG4}(a) shows the total time-dependent response $w(t\vert
\{\theta_{i}\})$ under three sets of parameters, $B=0.05,G=0$ (Case 1,
solid black curve), $B=0, G=0.05$ (Case 2, dashed green curve), and
$B=G=0.05$ (Case 3, solid red curve). We see that under
neuroadaptation of both parameters $\beta$ and $\Gamma^{\rm b}$ (Case
3), the transition to addiction occurs much faster (red curve), with a
shorter plateau in $W_{k}$ and a quicker drop in ${\rm RPE}_{k}$. In
general, a larger $\Gamma^{\rm b}$ relative to $\Gamma^{\rm a}$
depresses the time-dependent dynamical response $w(t\vert
\{\theta_{i}\})$ and the net response $W$. Larger $\alpha$ and $\beta$
lead to more transient responses that display less overlap between
intakes provided $T$ is fixed.  Smaller $\beta$ leads to longer
lasting b-processes that overlap across successive intakes.

In Fig.~\ref{FIG5} we explore the effects of varying the duration of
the a-process by setting $\beta_{1} = 0.5, \sigma = 0.05, \Gamma^{a} =
1, \Gamma^{b}_{1} = 0.1, B=0.05, G=0$ and changing $\alpha$. In Cases
4, (solid black curves), 5 (dashed green curves), and 6 (solid red
curves) we set $\alpha = 0.05, 0.2, 0.7$ to represent long-lasting,
intermediate, and short-lived a-processes, respectively.  As shown in
Fig.~\ref{FIG5}(a), a smaller $\alpha$ results in more overlap of
positive responses $w_{\rm a}$ and as a result, more positive overall
response $w(t\vert \theta_{i \leq k})$. Different values of $\alpha$
do not seem to appreciably change the number of intakes at which $w(t
\vert \theta_{i \leq k})$ becomes negative.  The evolution of
$\beta_{k}$ and $\Delta_{k}$ are nearly indistinguishable for all
three cases as shown in Fig.~\ref{FIG5}(b) and (c).
  
It is worth noting that, as shown in Fig.~\ref{FIG5}(c), the net
responses $W_{k}$ in Cases 4,5,6 reach long-lasting plateaus before
starting to decrease, between intakes $k=30$ and $k= 35$.  The
corresponding ${\rm RPE}$s shown in Fig.~\ref{FIG5}(d) fluctuate
around zero in all cases until relatively large intake numbers $k$ are
reached, indicating ``high-functioning'' users. Eventually however the
RPEs decrease and become negative as well.  However, the quickest
decent of the RPE$_{k}$ towards negative values is observed for the
longest lived a-process, Case 4 for $\alpha = 0.05$, whereas the most
stable ${\rm RPE}_{k}$ arises for the shorted lived a-processes, Case
6 for $\alpha = 0.7$, although the associated $W_{k}$ exhibits a
smaller amplitude. These results indicate that the sensitivity of the
responses, $W_{k}$, and ${\rm RPE}_{k}$ to changes in $\alpha$ are
nonlinear and involve a subtle interplay between the overlap of the a-
and b-processes the amplitude $\Gamma^{\rm b}$ of the b-process, and
the definition of the ${\rm RPE}$.

  \begin{figure*}[t]
  \centering
   \includegraphics[width=7in]{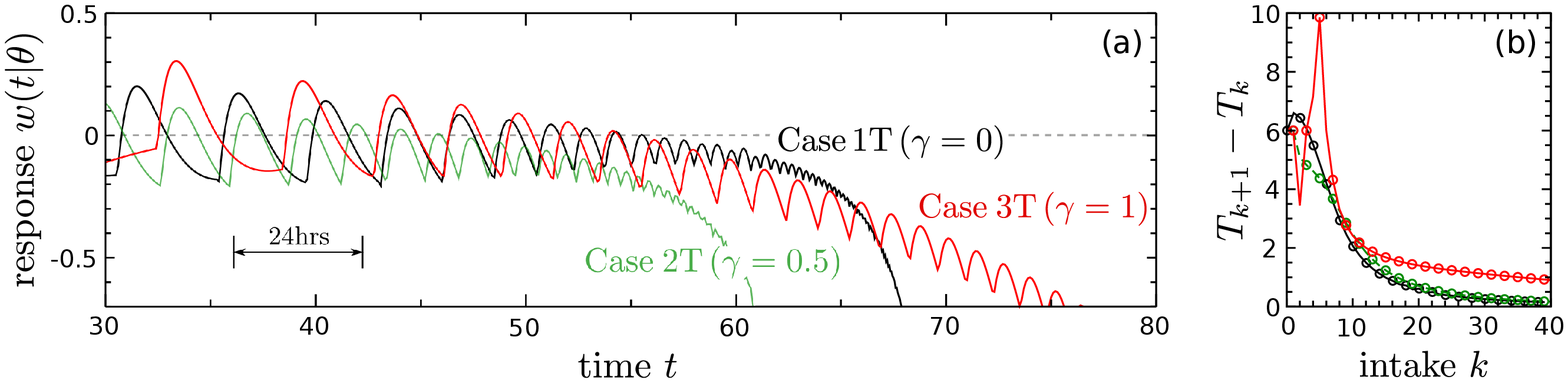}
  \caption{(a) Time-dependent response curves resulting from multiple
    drug intakes with varying $T_k$ are generated using $\alpha=0.5,
    \Gamma^{\rm a} = 1, B=G=0.01$ and $\beta_{1} = \Gamma^{\rm b}_{1}
    = 0.5$. Case 1T (solid black curve) assumes $\gamma_{k}=\gamma=0$
    and no memory before the last intake, while Cases 2T (solid black
    curve) and 3T (dashed red curve) assume intermediate and strong
    memory, $\gamma_{k}=\gamma=0.5$ and $\gamma_{k} = \gamma= 1$,
    respectively. Note that qualitatively, the decreasing trend is
    nonmonotonic in the memory $\gamma_{k}$. (b) Time separations
    $T_{k+1}-T_k$ between two consecutive intakes for the three
    cases.}
\label{FIG6}
\end{figure*}

The cases described above are illustrative of how different
user-specific parameters ($B,G,\sigma, \alpha$), and initial
conditions ($\beta_{1}, \Gamma^{\rm b}_1$) yield qualitatively
different paths to addiction.  Cases 1,2, and 3 reveal the effects of
higher neuroadaptive sensitivity (Case 3, $B=G=0.05$) whereby the
onset of addiction is dramatically faster.  Cases 4,5, and 6 compare
scenarios in which the trajectories of the neuroadaptive parameter
$\beta_{k}$ and intake dose $\Delta_{k}$ do not substantially
differ but can nonetheless lead to qualitative differences in the
magnitudes of the integrated response $W_{k}$, the drop-off point of
the ${\rm RPE}_{k}$, and the intake at which addiction occurs. The
yo-yo behavior of ${\rm RPE}_{k}$ is typically seen for users who are
allowed to adjust their doses through Eq.~\ref{RECURSION_DELTA}.

\subsection*{Evolution of intake timing}

We now consider the case where drug doses are equal for all intakes
$\Delta_{i}\equiv \Delta=1$, but the user-controlled intake times
$T_{i}$ do not define a periodic sequence.  Since $\Delta$ is
constant, the recursion relations~\ref{RECURSION_BG} are explicitly
solved by
$\beta_{i} =\beta_{1}(1-B\Delta)^{i-1}$ and $\Gamma^{\rm b}_{i} =
\Gamma^{\rm b}_{1}(1+G\Delta)^{i-1}$
under the assumption $B \Delta <1$.  These expressions represent
exponential decreases and increases in $\beta_{i}$ and
$\Gamma_{i}^{\rm b}$, respectively. Similar to how drug doses were
determined, we now assume that the user's decision of when to next
take drugs depends on the RPE defined in Eq.\,\ref{rpe}. Here, we set
$C_{k+1} = 0$ but keep the discount term $\gamma_{k-1} \leq 1$.  We
also assume that the next $k+1^{\rm th}$ intake occurs when RPE$_k(t)$
declines to the threshold value $R_{\rm c}$, representing the onset of
unpleasant effects after the high following intake $k$. Thus,
$T_{k+1}$ can be determined by the real root of
${\rm RPE}_{k}(T_{k+1}) =R_{\rm c}$ which, using the definition of 
 $\rm RPE_{k}$ reads
\begin{eqnarray}
\nonumber
W_{k}(T_{k+1}\vert \{\theta_{i \leq
  k},T_{i\leq k}\}) - 
& &\gamma_{k-1}W_{k-1}(T_{k}\vert \{\theta_{i \leq
  k-1},T_{i\leq k-1}\}) = R_{\rm c}. \\
\label{RECURSIONT}
\end{eqnarray}
This equation must be solved on the decreasing branch of the RPE$_{k}$
curve as the user takes the next dose to alleviate the decreasing net
response.  The user is ``initialized'' with daily intakes (of period
$T=6$ in non-dimensional units) until a real solution arises from
Eq.\,\ref{RECURSIONT}, indicating a user who adjusts their intake
timing to avoid unpleasant effects.  If at any time ${\rm RPE}_{k} =
R_{\rm c}$ again exhibits no real solution, we simply add $T$ to the
last intake time $T_k$ so that $T_{k+1} = T_k + T$.  In this case, the
user is satisfied with the effects of the $k^{\rm th}$ intake and
can return to his or her daily routine of drug consumption.
For concreteness, we set $\Delta_{k}=\Gamma^{\rm a}= 1, \alpha = 0.1,
B=G=0.01, \beta_{1}= \Gamma^{\rm b}_{1}=0.5$, and evaluate
$W_1(T_2=6)$.  We then generate $\{\beta_{2}, \Gamma^{\rm b}_{2}\}$
according to the exponential solutions to Eq.~\ref{RECURSION_BG}.  The
time of the third intake $T_{3}$ is then found by solving
$W_{2}(T_{3}\vert \{\theta_{i \leq 2},T_{i\leq 2}\}) -
\gamma_{1}W_{1}(T_{2}\vert \{\theta_{1},T_{1}=0\}) = R_{\rm c}$, and
so on.

\begin{figure*}[t]
  \centering
  \includegraphics[width=7in]{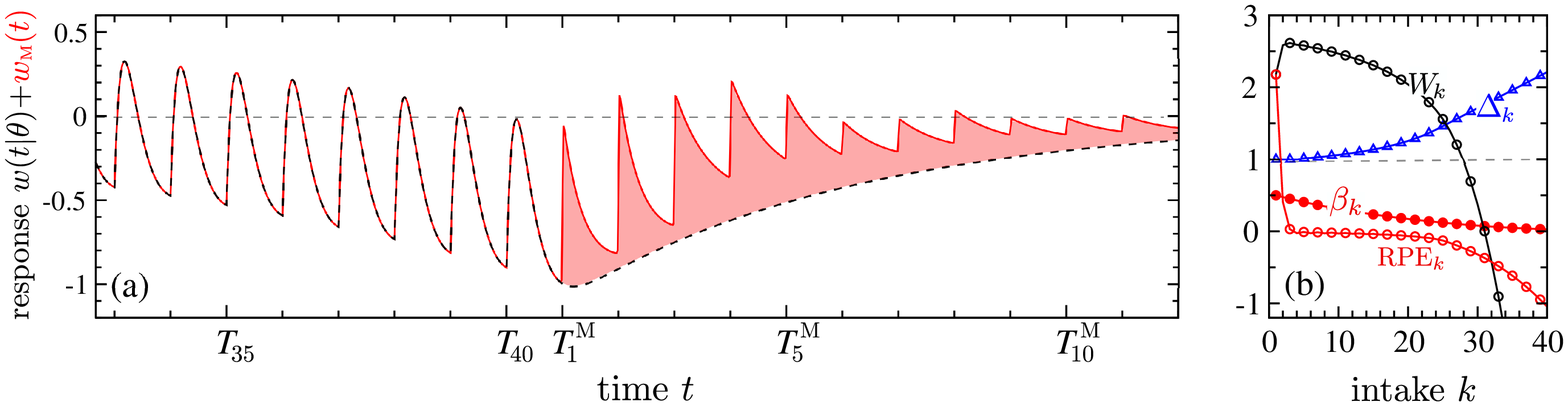}
  \caption{(a) Time-dependent response $w(t\vert \theta)$ (black
    dashed curve) with the superimposed methadone contribution $w_{\rm
      M}(t)$ (red solid curve). The time-dependent response in the
    absence of methadone returns to the baseline over a timescale
    $\sim 1/\beta_{40}$ producing unpleasant withdrawal symptoms
    during this time.  Methadone treatment ($+w_{\rm M}(t)$) adds to
    the response reducing the negative effects of b-processes by an
    amount indicated by the red shaded area. (b) $\beta_{k}$,
    $\Delta_{k}$, $W_{k}$, and ${\rm RPE}_{k}$ associated with the
    drug sequence prior to the administration of methadone.}
\label{FIG7}
\end{figure*}

In Fig.\,\ref{FIG6} we consider three scenarios representing different levels of memory of
the previous intake reflected by different values of $\gamma_{k-1}$ in
Eq.~\ref{rpe}.  In Case 1T (solid black curve), we set $\gamma_{k-1} =
0$ to describe a user who does not remember the response from any
previous dose and only uses the current net response $W_{k}(t)$ to
determine the next intake at time $T_{k+1}$. As shown in
Fig.~\ref{FIG6}(a), the intakes become successively more frequent
giving rise to a sharp decline in the dynamic response $w(t \vert
\theta_{i \leq k})$ after about $t \approx 65$, about a week if $T=6$
corresponds to 24 hrs. In Case 2T (dashed green curve), we set
$\gamma_{k}=0.5$ to describe a user who weights the net response of
the previous intake, $W_{k-1}$ half as much as that relative to the
current intake, $W_k$. In this case, full addiction occurs at intake
$k=4$ (not explicitly shown) at time $t\approx 20$, about three
days. In Case 3T (solid red curve), $\gamma_{k}=1$ and the user fully
remembers the response associated with the previous dose in his or her
determination of the next intake. In this case the response decreases
more slowly than in Cases 1T and 2T. The decreases occurs later than
when $\gamma=0.5$ but earlier than when $\gamma_{k}=0$.  In
Fig.~\ref{FIG6}(b) we plot $T_{k+1} - T_k$ for all three cases which
show subtle differences in timing associated with the three
qualitatively different cases. In Case 3T, the slower decrease in
successive $T_{k}$s for large $k$ results from the slower drop-off of
$w(t\vert \theta_{i \leq k})$ at long times.

Since by construction, ${\rm RPE}_{k}(T_{k+1}) = R_{\rm c}<0$ for all
$k$, addiction is reached at intake $k^{*}$ such that $W_{k^*}<0$.  If
the first $i\leq i^{*}$ intakes are taken at fixed times
$T_{i}=(i-1)T$ because Eq.~\ref{RECURSIONT} has not yet generated a
real root, the first intake for which $W_{k^{*}}<0$ occurs at $k^{*}
\approx i^{*}+j^{*}$ where $j^{*}$ is found by the lowest integer $j$
such that

\begin{equation}
\sum_{\ell=1}^{j-2}\gamma^{-\ell} > -{W_{i^{*}}\over R_{\rm c}}
\end{equation}
for constant $\gamma_{k} = \gamma$. A related form can be easily
derived when $\gamma_{k}$ depends on $k$. For Cases 1T, 2T, and 3T
shown here, $k^{*} \approx 2,4,10$, respectively.  In
general, we find that the iteration of Eq.~\ref{RECURSIONT} continues
until either the inter-intake times $T_{k+1}-T_{k} \to 0$, or no
positive real root can be found, indicating an RPE that is permanently
below the threshold value $R_{\rm c}$ and that the user's expectation
can never be met. The loss of the root is more likely to arise when
$\beta_{k}$ and $\Gamma^{\rm b}_{1}$ are small but always occurs after
$W<0$ and ${\rm RPE}<0$ (addiction).

The above examples show that the protracted use of drugs leads to
neuroadaptive decreases in $\beta$ and more slowly decaying
b-processes. In the limiting case $\beta\to 0$, the user appears to be
in an allostatic state, with near-permanently damaged brain circuits
and altered reward response baseline levels. Note that a true
allostatic state can be defined within our model by replacing $-\beta
w_{\rm b}(t)$ in Eq.~\ref{wb} with a term such as $-\beta(w_{\rm b}(t)
+ w_{\infty})$.  The infinite time response would then relax to
$w_{\rm b}(\infty) \to -w_{\infty}$. This new baseline level may
itself evolve after repeated intakes via neuroadaptive processes
similar to those represented by Eqs.~\ref{RECURSION_BG}.

\subsection*{Mitigation through agonist intervention}

Our model provides a framework to study detoxification strategies
where dosing of substitutes, such as methadone in the case of heroin
addiction, can be calibrated to alleviate withdrawal symptoms without
producing euphoric effects \cite{Jackson1973}.  We assume an
``auxiliary" drug, such as methadone, operates on a related, but
different, pathway of the brain reward system relative to the ones
stimulating the a- and b-processes described in Eqs.\,\ref{wa} and
\ref{wb}. This auxiliary drug may generate a separate reward response
which itself may evolve according to neuroadaptation or interactions
with other neural networks.  We denote the additional reward response
$w_{\rm M}(t)$ so that, within the context of our model, the overall
user perception is given by the sum $w_{\rm a} + w_{\rm b} + w_{\rm
  M}$.  Positive values $w_{\rm M}>0$ shift the overall response
towards the homeostatic baseline, reducing withdrawal symptoms. If
neurocircuits are not permanently damaged, our results imply that an
ideal treatment consists of applying a large enough $w_{\rm M} > 0$
that mitigates the negative response $w_{\rm b}$ over a timescale
$\sim 1/\beta_k$, where $\beta_k$ is the value of the b-process decay
rate at the time of last intake.

To be concrete, we model a hypothetical heroin addiction via
an intake sequence associated with $\alpha = 0.3,\beta_{1}=0.5,
\Gamma^{\rm a}=1, \Gamma^{\rm b}=0.1, B=0.05, G=0, \sigma=0.05, R_{\rm
  c}=-0.05$, and $T=6$. In Fig.~\ref{FIG7}, we show the response
starting at $t\approx 200$ corresponding to approximately intake $33$
(in this example, full addiction occurred at intake $k^{*}=32$). We
assume the user subsequently ceases heroin consumption at intake
$k=40$ where $\beta_{40} \approx 0.02$ and $\Delta_{40} \approx
2.3$. The user is then assumed to start a methadone maintenance
treatment following a protocol of $\Delta_k^{\rm M}$ doses at
prescribed times $T_k^{\rm M}$. We model the methadone response as

\begin{equation}
\label{w_meth}
    w_{\rm{M}}(T_{k}^{\rm M} < t < T_{k+1}^{\rm M}\vert
\Delta^{\rm M}_{i\leq k},\delta^{\rm M}_{i\leq k},
T^{\rm M}_{i\leq k})= \sum_{i=1}^{k}\Delta_i^{\rm M}e^{-\delta_i^{\rm M}(t-T_{i}^{\rm M})}.
\end{equation}
where the dimensionless decay rates $\delta_i^{\rm M}$ are measured
relative to the overall dopamine clearance rate $\delta$ discussed in
Eq.~\ref{activity}. Eq.~\ref{w_meth} is a succinct representation of
the user perception of methadone; $1/\delta_i^{\rm M}$ represents an
effective lifetime that depends on the decay of methadone in the body
and of the effects of the associated reward. A more complex model can
be developed along the lines of Eq.~\ref{wa}.
The lifetime of methadone in the body changes as treatment progresses,
and ranges from initial values of $10-20$hrs to $25-30$hrs in the
maintenance phase. In clinical settings, $\Delta_k^{\rm M}$ also
typically increases \cite{Who2009}; for example, the first methadone
doses range between $10-30$mg while later doses are increased to about
$60-120$mg. Methadone dosage can also depend on the user's history of
opioid use.

In our model we apply 11 daily methadone doses, with the drug
administered at periodic intervals of $T=6$ starting at time $t=246$,
a day after the last $k=40$ heroin intake at $t=240$. We assume that
the methadone doses follow the sequence $\Delta^{\rm M}_{i} =
\{1,1,0.8,0.6,0.4,0.2,0.2,0.2,0.1,0.1,0.1\}$ and that neuroadaptation
increases the methadone timescale from about 10hrs to 27hrs
leading to $\delta_{i}^{\rm M}=
\{0.4,0.4,0.2,0.2,0.2,0.15,0.15,0.15,0.15,0.15,0.15\}$.  Figure
\ref{FIG7} shows the methadone-induced response $w_{\rm M}(t)$ (red
curve) added to the drug-induced response (black dashed curve).

Note that without methadone treatment, once heroin consumption ceases
after intake $k=40$, the time-dependent reward response (black dashed
curve) resembles an allostatic load which returns to the baseline over
a long timescale $\sim 1/\beta_{40}\approx 50$, over a week.  The
methadone-derived response $w_{\rm M}(t)$ (red curve) alleviates much
of the negative b-process and associated withdrawal symptoms.  The net
time-integrated reduction in withdrawal symptoms is represented by the
red shaded area between the $w_{\rm a} + w_{\rm b}$ and the $w_{\rm a}
+ w_{\rm b} + w_{\rm M}$ curves as shown in Fig.~\ref{FIG7}.

Although methadone is used to treat addiction, it is an opioid agonist
and can itself induce addiction through $w_{\rm M}$ which may also
trigger its own b-processes.  This is especially true if methadone is
taken in an uncontrolled manner, and may explain why often suboptimal
doses are administered \cite{Szalavitz2017}.  Thus, control of $w_{\rm
M}(t)$ is crucial in using methadone as a treatment.  An ideal
protocol would calibrate doses and timing to alleviate the negative
response as much as possible, but would also prevent the induction of
methadone-associated b-processes, or other interactions with addictive
pathways. One can also explore the consequences of irregular methadone
intakes or nonadherence to specific detoxification
protocols \cite{Lawley2022}.

\section*{Discussion and Conclusions}

We constructed a quantitative framework for the evolution of drug
addiction based on concepts from IST, OPT and where drug dosages
depend on the RPE. Our goal was to develop an explicit model that
incorporates these key ingredients in a simple and clear way, without
invoking a large number of parameters. Although many models that
include action choice have been developed \cite{CHANGEUX,book_adapt},
our work assumes only one dominant action (drug taking) that emerges
from the background response to all other routine rewards.  We are
thus assuming that the response to these ``normal'' rewards has
already been subtracted from the drug-specific response $w_{\rm
a}+w_{\rm b}$. A much richer stochastic model can be developed by
considering fluctuating responses from routine rewards.

In our model, repeated intakes lead to overall negative reward
responses due to neuroadaptive processes that lessen drug-induced
pleasurable effects.  To counterbalance this shift, the user actively
seeks higher rewards by increasing drug dose, intake frequency, or
both. These behaviors create a feedback loop that induce further
neuroadaptive changes and that eventually lead to an addicted state.
Our model captures the well-known phenomenon of tolerance by allowing
expectations to increase after a drug intake, which in turn leads the
user to increase the dosage as an attempt to meet the new expectation
level.  Mathematically this is represented by allowing the RPE to fall
below a critical threshold value.  Our model can also explain the
increased frequency of drug intaking by dictating that the user takes
a new dose once the RPE reaches the critical threshold value.  A more
realistic description would define an objective function that allows
the user to both increase drug dosage and to take it more often.
 
How addiction unfolds depends on the specific physiology and
neuroadaptive response of the user.  Within our simple mathematical
model, the path to addiction depends on the sequence of representative
parameters that change with each drug intake $i$.  These parameters
represent neuroadaptive characteristics such as
$\{\alpha_i,\Gamma^{\rm a}_i,\beta_i,\Gamma_i^{\rm b}\}$ that appear
in Eqs.\,\ref{wa} and \ref{wb} as well as user-controlled dosing
$\Delta_{i}$ and timing $T_{i}$ that dictate the evolution of the RPE.
In our analyses, we fixed $\alpha_i$ and $\Gamma^{\rm a}_i$ and
proposed simple recursion relations for $\Delta_i,\beta_i$, and
$\Gamma^{\rm b}_i$ that evolve consistently with OPT.  Specifically,
this scheme represents b-processes becoming more prominent as drug
addiction unfolds.  
If a user is genetically predisposed to addiction 
or if the drug is highly addictive, as in the case of 
methamphetamines, the parameters $R_{\rm c}$ and $\sigma$,
and $B$ and $G$ that drive the evolution
  of the b-process will be larger and the number of intakes necessary
to reach the addicted state will be few. For more resistant users
and/or slowly addictive substances such as cannabinoids, $R_{\rm c},
\sigma, B$, and/or $G$ will be smaller, leading to a more drawn-out
addiction process that includes damped oscillatory progression of the
RPE. We also find that reaching the addicted state
  will require less intakes if the onset value of $\Gamma_{\rm b}/\beta >
  1$, implying an initially strong and persistent b-process. These
  results allow us to predict that the most at risk users are those
  who are most reactive to changes in the b-process and (assuming that
  the brain processes all rewards through the same pathway) those who
  manifest elevated b-process responses even prior to drug intaking.  Although there are not many studies
  connecting personality traits with addiction \cite{Widiger2017}, our
  finding is consistent with reports of neurotic individuals being
  among the most at risk for drug addiction \cite{Turiano2012}. One of
  the main hallmarks of neuroticism in fact is for negative affects,
  such as the ones expressed by the b-process, to be more pronounced
  \cite{Watson1988,Luhmann2018}.

One simplification of our analysis is that we considered either
variable doses $\Delta_i$ administered at periodic intervals $T$, or
constant doses $\Delta$ taken at non-uniformly spaced timings $T_i$. A
more comprehensive study would allow for the RPE to dictate both
dosages $\Delta_i$ and timings $T_i$ as a function of expectations
built on previous drug intakes, without fixing either a priori.

A number of refinements to our model can be straightforwardly
incorporated. For example, instead of a sequential response to drug
intake, where $w_{\rm b}$ is triggered by $w_{\rm a}$, one could
consider a parallel response where the drug-induced dopamine surge
triggers both $w_{\rm a}$ and $w_{\rm b}$.  Similarly we could
consider a networked response, with several pleasurable and aversive
neuronal centers being activated and/or stimulating one another.
Alternatively, one could consider a multicomponent reward response
that depends on neuronal sets differentially activated by multiple
drugs.  To study this case, one would need to derive a single-output
reward response from a high-dimensional multi-drug input.  If the
multiple drugs lead to neuroadaptive changes in the relaxation rates
$\alpha$ and $\beta$, their effects on the rewards $w_{\rm a}$ and
$w_{\rm b}$ would be multiplicative. Different drugs may have
different \textit{in vivo} clearance rates and drive dopamine release
with different durations leading to different dopamine residence times
$1/\delta^{(j)}$.  They may also activate different sets of neurons
that contribute to the a- and b-processes $w_{\rm a}$ and $w_{\rm b}$
additively through the weights $\Gamma^{\rm a}$ and $\Gamma^{\rm b}$.
Thus, multiple drugs potentially administered at different times, can
contribute to the overall response both additively and
multiplicatively, leading to rich dynamical behavior of the brain
reward system. The inclusion of broader action classes (or
``policies'') can also be incorporated using a more formal framework
from reinforced learning \cite{book}.

Another possible approach would be to include continuous-time
evolution of the parameters $\{\alpha, \Gamma^{\rm a},\beta,
\Gamma^{\rm b}\}$ or to include more realistic forms for the RPE such
as a convolution of a memory kernel with $w_{\rm a} + w_{\rm b}$ as
motivated from data \cite{CHANGEUX,Redish2004}. More complex nonlinear
evolution of parameters could also be considered which could give rise
to sharper transitions into an addictive
state \cite{CHANGEUX} and to chaotic behaviors \cite{Li1975, Warren2003}.
Sharper transitions would be partially
mitigated by an RPE definition where current rewards are compared with
averages over past periods.  Although we assumed a well-defined
``deterministic'' behavioral rule for changing drug dose and intake
timing, prolonged drug use can lead to dysfunction in decision-making
and unpredictable and random behavioral changes \cite{Sweis2018},
justifying nonlinear dynamics and/or stochasticity in the definition
of an effective RPE. Note that this stochasticity applied to the RPE
would be different from adding noise to $D(t)$ and treating
Eqs.~\ref{wa} and \ref{wb} stochastically (\textit{e.g}, as a Langevin
equation).  One can also examine in more detail the intake-dependent
additive cue in Eq.\,\ref{rpe} to predict how the RPE changes when
moving from a controlled drug-taking environment (where cues such as
location, paraphernalia, and accessibility are constant) to a more
random one (where cues may vary in time and across intakes). Cues can
also trigger dopamine releases without any actual drug intaking
\cite{Volkow2011} and can lead to relapses after long periods of abstinence
when the memory of previous intakes has subsided. Finally,
our model can be generalized to other forms of chemical or behavioral
addictions, such as alcoholism, gambling, or social-media addiction.

\begin{acknowledgments}
The authors thank Xiaoou Cheng for insightful comments. This research
was supported by the Army Research Office through grant
W911NF-18-1-0345, the NIH through grant R01HL146552 (TC), and the NSF
through grant DMS-1814090 (MD).
\end{acknowledgments}

\section*{Data Availability Statement}
Data available on request from the authors.  The data that support the
findings of this study are available from the corresponding author
upon reasonable request.

\bibliography{refs}

\end{document}